\documentclass[a4paper,12pt]{article}
\usepackage{amssymb,amsmath}
\usepackage{bm}%
\usepackage{stmaryrd}
\usepackage{slashed}
\usepackage[pdftex]{hyperref}
\hypersetup{
	colorlinks=true,
	citecolor=blue,
	linkcolor=blue,
	urlcolor=orange,
}
\usepackage{cite}
\usepackage{comment}
\usepackage{xcolor}

\def\Z2{\mathbb{Z}_2^2}

\def\nn{\nonumber}

\def\g{\mathfrak{g}}

\begin{document}

\title{$\Z2$-graded supersymmetry via superfield on minimal $\Z2$-superspace}

\author{N. Aizawa, Ren Ito, Toshiya Tanaka
	\\[10pt]
	Department of Physics, Graduate School of Science, \\ Osaka Metropolitan University, \\
	Nakamozu Campus, Sakai, Osaka 599-8531, Japan}

\maketitle
\thispagestyle{empty}

\vfill
\begin{abstract}
A superfield formalism for the minimal $\Z2$-graded version of supersymmetry is developed. 
This is done by using the recently introduced definition of integration on the minimal $\Z2$-superspace. 
It is shown that one may construct $\Z2$-supersymmetric action by the procedure similar to the standard supersymmetry. 
However, the Lagrangian obtained has very general interaction terms, which give rise to a $\Z2$-graded extension of many known theories defined in two-dimensional spacetime. 
As an illustration, we will give a $\Z2$-extension of the sine-Gordon model different from the one already discussed in the literature. 
\end{abstract}

%%%%%%%%%%%%%%%%%%%%%%%%%%%%%%%%%%%%%%%%%%%%%%%%%%%%%%%%%%%%%%%%%%%%%%%%%%%%%%%%%%%%%%
\clearpage
\setcounter{page}{1}

\section{Introduction}

Consider a vector space graded by an Abelian group $\mathcal{A}.$ 
One may define an algebra on this space by introducing a generalized Lie bracket which respects the grading structure \cite{Ree} (see also \cite{RW1,RW2,sch1}). If the vector space has no grading, the defined algebra is nothing but a Lie algebra. For $\mathcal{A} = \mathbb{Z}_2, $ it coincides a Lie superalgebra whose defining relations are given in terms of commutators and anticommutators. If $ \mathcal{A} = \mathbb{Z}_2^n := \mathbb{Z}_2 \times \mathbb{Z}_2 \times \cdots \mathbb{Z}_2 $ ($n$ times), the algebra is also defined in terms of (anti)commutators so that a natural generalization of Lie superalgebras. 

These extensions of Lie superalgebras have opened up  new areas of research of great interest in both physics and mathematics. 
In physics, symmetries generated by $\Z2$-graded Lie superalgebras have been found in various places, e.g., de Sitter SUGRA \cite{Vas}, nuclear quasi-spin \cite{jyw}, parastatistics \cite{tol}, non-relativistic Dirac equation \cite{AKTT1,AKTT2} and so on. 
This implies that the $\mathbb{Z}_2^2$-graded Lie superalgebras are not unusual in physics, but offer a new insight into the analysis of physical systems based on symmetries, which is one of the  motivations to study their structure and representations. Structure and representation theories of $\mathbb{Z}_2^n$-graded Lie superalgebras have been studied continuously since their introduction \cite{sch3,SchZha,Sil,ChSiVO,SigSil,PionSil,CART,MohSal,NAJS,NAPIJS,StoVDJ,Meyer,IsStvdJ,NAPSIJSinf,NAKA,KTclassification,Meyer2,NeliJoris,NeliJoris2,LuTan,FaFaJ}.  
For physical applications, however, more needs to be done. 

It is natural to consider $\mathbb{Z}_2^n$-graded extensions of supersymmetry by generalizing the super-Poincar\'e algebra. The generalization of a given superalgebra to $\mathbb{Z}_2^n$-setting is, in general, not unique.  There are some proposals of $\mathbb{Z}_2^n$-graded extensions of the spacetime supersymmetry \cite{zhe,LR,Toro1,Toro2,tol2,Bruce}. Among others, the $\mathbb{Z}_2^n$-graded extension of  super-Poincar\'e algebra introduced by Bruce \cite{Bruce}   attracts much interest and $\mathbb{Z}_2^n$-graded version of quantum mechanics, sigma model, etc. has been discussed based on the extension \cite{BruDup,AAD,AAd2,AKTcl,AKTqu,DoiAi1,DoiAi2,AiDoi,brusigma,bruSG,AIKT}.  
The $\Z2$-graded supersymmetric quantum mechanics \cite{BruDup} ($\Z2$-SQM) have been shown to be nontrivial since there exist observables that can detect the difference between $\Z2$-SQM and the standard one \cite{Topp,Topp2}. 
The $\mathbb{Z}_2^n$-supersymmetry introduces a new type of para-particles which are different from the traditional ones.  
Recently, the traditional para-particles are simulated by using a trapped ion \cite{ParaP}, so we expect that $\mathbb{Z}_2^n$-para-particles will also be realized in the laboratory. Therefore, it would be important to study para-particles from different perspectives.

For a deeper understanding of the symmetries generated by the $\mathbb{Z}_2^n$-graded algebras, it would be helpful to analyse  more models with such symmetries. 
For this, we need to have more models with the symmetries. 
One possible way to build models will be to extend the superfield formalism and build classical systems with the $\mathbb{Z}_2^n$-graded supersymmetry. 
Recall that in ordinary supersymmetry the superfield is a function on a superspace and the classical action is given by integrating a Lagrangian, which is a function of the superfields, over the superspace. 

The $\mathbb{Z}_2^n$-graded extension of superspace, superfields can be done without any difficulty. 
Differential calculus on the superspace has also been established \cite{CoKPo}. 
However, integration is only considered  for the simplest (minimal) $\Z2$-superspace and has not yet been established even for the simplest case. 
In fact, two different definitions of the integration on the minimal $\Z2$-superspace are proposed \cite{Pz2nint,PonSch} and there is no consensus in the mathematical community. 
These studies are a part of $\mathbb{Z}_2^n$-supergeometry which is one of the  fields of research  opened up by the $\mathbb{Z}_2^n$-graded superalgebras \cite{Covolo,CGP1,CGP2,CGP3,CGP4,BruIbar,BruPon,BruIbarPonc,BruGraRiemann,BruGrabow,CoKwPon,BruIbarPon2,BruGrabow2} (\cite{PonSch} contains a concise review of this field). 

In the recent work \cite{NARI}, two of the present authors studied the integration on the minimal $\Z2$-superspace in detail and introduced a new definition of the integral. 
The purpose of the present paper is to develop the $\Z2$-graded version of the superfield formalism on the minimal $\Z2$-superspace by using the new integral. 
We shall see that the new integral allows us to construct $\Z2$-supersymmetric classical actions  by a  procedure similar to the standard supersymmetry. 
The action obtained has an additional symmetry corresponding to the $\Z2$-graded Lorentz transformation in the minimal $\Z2$-superspace. 
Furthermore, it has very general interaction terms, which give a $\Z2$-graded extension of many known theories defined in two-dimensional spacetime. 
As an illustration, we will give a $\Z2$-extension of the sine-Gordon model different from the one already discussed in \cite{bruSG}.

The success of the $\Z2$-superfield formalism reflects two properties of the new integral and shows its validity.
The properties are: (i) no restriction is imposed on the integrand if its components are compactly
supported, or at least if they vanish fast enough at infinity. 
This is a contrast to the integration in \cite{Pz2nint,PonSch} and allows us to set physically relevant Lagrangians. 
(ii) the integral emerges a spatial coordinate. More precisely, the minimal $\Z2$-superspace has three noncommutative coordinates and the integration with respect to one of them is defined by a mapping to the ordinary integration of a real number. The real number plays the role of the emergent space coordinate. Together with the time coordinate of the minimal $\Z2$-superspace we have a Lagrangian defined on two-dimensional spacetime.

The plan of this paper is as follow: 
\S \ref{SEC:Minimal} is a preliminary where the basics of the minimal $\Z2$-superspace, that is, definition, transformations and differential calculus are summarized. We also present the definition of integration used throughout the paper. 
\S \ref{SEC:SupefieldFromalization} describes how to construct the invariant action. 
In \S \ref{SubSEC:construction}, starting from the definition of the $\Z2$-superfield and covariant derivative, we give a Lagrangian covariant under the $\Z2$-Lorentz transformation in addition to the $\Z2$-SUSY transformation. 
Then, integration over the minimal $\Z2$-superspace gives us the action in terms of physical fields. 
We explicitly present  the equations of motion and the conserved currents associated with all the symmetries  in \S \ref{SubSEC:EoM}. 
The action obtained in \S \ref{SEC:SupefieldFromalization} is quite general, it allows various interactions. 
We give two examples of the interaction in \S \ref{SEC:examples}. 
One is the square of the $\Z2$-superfield which makes bosonic fields massive.  
The other is the cosine of the $\Z2$-superfield which gives a $\Z2$-graded extension of the sine-Gordon equation. 
We summarize the results and give some comments on future direction in \S \ref{SEC:Conclusions}.

%%%%%%%%%%%%%%%%%%%%%%%%%%%%%%%%%%%%%%%%%%%%%%%%%%%%%%%%%%%%%%%%%%%%%%%%%%%%%%%%%%%%%%
%
\section{Minimal $\Z2$-superspace and calculus} \label{SEC:Minimal}
\setcounter{equation}{0}

Let us first recall the definition of $\Z2$-graded Lie superalgebras \cite{RW1,RW2}. 
A $\Z2$-graded vector space  (over $\mathbb{R}$ or $ \mathbb{C}$) is the direct sum of homogeneous vector subspaces labeled by an element of $\Z2$:
\[
\g = \g_{(0,0)} \oplus \g_{(1,1)} \oplus \g_{(1,0)} \oplus \g_{(0,1)}.
\]
An element of $ \g_{\vec{a}}$  is said to have the $\Z2$-\textit{degree} $ \vec{a}  \in \Z2.$ 
We define the $\Z2$-Lie bracket by
\begin{equation}
	\llbracket X, Y \rrbracket = XY - (-1)^{\vec{a}\cdot\vec{b}} Y X,
	\quad
	X \in \g_{\vec{a}}, \ Y \in \g_{\vec{b}}
\end{equation}
where $ \vec{a}\cdot\vec{b} $ is the standard scalar product of two dimensional vectors. 
Namely, the $\Z2$-Lie bracket is the commutator (anti-commutator) for $ \vec{a}\cdot\vec{b} $ is even (odd). 
A $\Z2$-graded vector space is said to be a $\Z2$-graded Lie superalgebra if 
$ \llbracket X, Y \rrbracket \in \g_{\vec{a}+\vec{b}} $ and the Jacobi identity is satisfied:
\[
\llbracket X, \llbracket Y, Z \rrbracket \rrbracket
= \llbracket \llbracket X, Y \rrbracket, Z \rrbracket
+ (-1)^{\vec{a}\cdot \vec{b}} \llbracket Y, \llbracket X, Z \rrbracket \rrbracket.
\]
If $ \llbracket X, Y \rrbracket = 0, $ we say that $ X $ and $Y$ are $\Z2$-\textit{commutative}. 
We also define the even and odd subspaces of $\g$ by $ \g_{(0,0)} \oplus \g_{(1,1)} $ and $\g_{(1,0)} \oplus \g_{(0,1)},$ respectively.  

The $\Z2$-graded Lie superalgebra considered in this work is the one-dimensional $\Z2$-graded supersymmetric algebra ($\Z2$-SUSY algebra) introduced in \cite{Bruce} which is an $\Z2$-graded extension of the super-Poincar\'e algebra in one-dimensional spacetime. 
Each subspace of the $\Z2$-SUSY algebra is one-dimensional
\begin{equation}
	H \in \g_{(0,0)}, \quad Z \in \g_{(1,1)}, \quad Q_{10} \in \g_{(1,0)}, \quad Q_{01} \in \g_{(0,1)}
\end{equation}
and the non-vanishing $\Z2$-Lie brackets  in terms of commutator or anticommutator are given by
\begin{align}
	\{Q_{10},Q_{10}\}=&\{Q_{01},Q_{01}\}=2H
	,\qquad
	[Q_{10},Q_{01}]=iZ. 
	\label{SUSYalg}
\end{align}
As in the standard supersymmetric theory, the space dual to the $\Z2$-SUSY algebra is $\Z2$-commutative and identified with a space with $\Z2$-grading \cite{AIKT}:
\begin{equation}
	g = \exp (itH-\theta_{10}Q_{10}-\theta_{01}Q_{01}+izZ). \label{Gelement}
\end{equation}
The $\Z2$-superspace $(t,z,\theta_{10},\theta_{01})$ is minimal in the sense there exits only one elements in each graded subspaces. The coordinate $t$ of degree $(0,0) $ may be identified with time, $ \theta_{10}, \theta_{01}$ are nilpotent fermionic coordinate, but they commute each other. There exists extra bosonic degree $(1,1)$ coordinate $z$ which anticommutes with $\theta_{10}, \theta_{01}$.

The operation of star conjugation $\ast$, which allows to define the hermitian operators,  is defined for graded elements $a,b$ and $\lambda\in {\mathbb C}$ to satisfy
\begin{equation}
 (ab)^\ast = b^\ast a^\ast, \quad (a^\ast)^\ast =a,\quad (\lambda a)^\ast =\lambda^\ast a^\ast,
\end{equation}
where $\lambda^\ast $ denotes the complex conjugation of $\lambda$.
Throughout the paper we assume that
\begin{alignat}{4}
	H^\ast &=H, & \quad Q_{10}^\ast& =Q_{10}, & \quad Q_{01}^\ast&=Q_{01}, & \quad Z^\ast = Z,
	\nn \\
   t^\ast &= t, & \theta_{10}^\ast &=\theta_{10}, & \theta_{01}^\ast&=\theta_{01}, & z^\ast=z. 
   \label{reality}	
\end{alignat}
It follows that $g$ is unitary: $ g^{-1} = g^\ast.$ 

The left action of a group element 
\begin{equation}
	g' = g_\epsilon \cdot g,\qquad {\textrm{with}} \qquad 
	g_\epsilon = \exp (i\epsilon_{00}H-\epsilon_{10}Q_{10}-\epsilon_{01}Q_{01}+i\epsilon_{11}Z)
\end{equation}
for the infinitesimal $ \epsilon_{\vec{a}}\ (\epsilon_{\vec{a}}^\ast = \epsilon_{\vec{a}})$  parameters with $\Z2$-degree induces a $\Z2$-SUSY transformation of the $\Z2$-superspace. 
\begin{equation}
	g = g(X) \ \to \ g'(X') = g(X+\delta X)
\end{equation}
where $ X \in \{ t, z, \theta_{10}, \theta_{01} \}$. More explicitly, we have the following transformations:
\begin{alignat}{2}
	\delta t &= \epsilon_{00}+ i\epsilon_{10}\theta_{10}+i\epsilon_{01}\theta_{01}, &\qquad \delta z &= \epsilon_{11}+\frac{1}{2}(\epsilon_{10}\theta_{01}-\epsilon_{01}\theta_{10}),\nonumber\\
	\delta \theta_{10} &= \epsilon_{10},  & \delta\theta_{01} &=\epsilon_{01}. 
	\label{CoordinateSUSYTransf}
\end{alignat}
The generators of these transformations are given by
\begin{align}
	\delta X &= (-i\epsilon_{00} H - i\epsilon_{11} Z + \epsilon_{10} Q_{10} + \epsilon_{01} Q_{01}) X,
	\nn \\
	H &= i\partial_t, \qquad Z = i\partial_z,
	\nn \\
	 Q_{10}=&\partial_{10}+i\theta_{10}\partial_t+\frac{1}{2}\theta_{01}\partial_z, 
	 \qquad 
	  Q_{01}=\partial_{01}+i\theta_{01}\partial_t-\frac{1}{2}\theta_{10}\partial_z
\end{align}
where we use the same notation for the generators and the algebra. 
The derivative with respect to the coordinates is defined as usual: 
\begin{equation}
	\partial_X X^k = k X^{k-1}.
\end{equation}  
We remark that the derivatives have the same $\Z2$-degree as the coordinates so that they are $\Z2$-commutative among themselves. 

In this work, we consider an additional transformation generated by
\begin{equation}
	L_{11} = -2i z \partial_t -\frac{i}{2}t\partial_z + \frac{1}{2} (\theta_{01} \partial_{10}-\theta_{10}\partial_{01})
\end{equation}
which causes, according to  
$ \delta_L X = -i\epsilon_L L_{11} X,$  the coordinate transformation
\begin{align}
	\delta_L t = -2\epsilon_L z, \quad 
	\delta_L z = -\frac{\epsilon_L}{2} t, \quad 
	\delta_L \theta_{10} = -\frac{i}{2}\epsilon_L \theta_{01}, \quad 
	\delta_L \theta_{01} = \frac{i}{2}\epsilon_L \theta_{10}
	\label{CoordinateLorentzTransf}
\end{align}
where $ \deg{\epsilon_L} = (1,1)$. 
This transformation mixes the bosonic coordinates, thus it may be interpreted as degree $(1,1)$ Lorentz transformation. 
The generator $ L_{11} $ together with the $\Z2$-SUSY algebra forms an five-dimensional $\Z2$-graded Lie superalgebra which we denote simply by $\g$. 
The additional relations are given by
\begin{alignat}{2}
	[L_{11}, H] &= \frac{i}{2}Z, & [L_{11}, Z] &= 2iH,
	\nonumber \\
	\{L_{11}, Q_{10}\} &=-\frac{1}{2}Q_{01}, &\quad \{ L_{11}, Q_{01}\} &= \frac{1}{2}Q_{10}
	\label{SUSY-Lorentz}
\end{alignat}

Functions on the minimal $\Z2$-superspace $(t,z,\theta_{10},\theta_{01}) $ are defined as a formal power series of $ z, \theta_{10} $ and $ \theta_{01}.$ By taking into account the nilpotency of $\theta_{10}, \theta_{01}$ one may expand
\begin{align}
	\Phi(t,z,\theta_{10},\theta_{01}) 
	&= \sum_{k=0}^{\infty}  f_{k00}(t) z^k + \theta_{10}\left( \sum_{k=0}^{\infty}  f_{k10}(t) z^k \right) 
	\nn \\
	&+ \theta_{01} \left( \sum_{k=0}^{\infty}  f_{k01}(t) z^k \right)  +  \theta_{10} \theta_{01}\left( \sum_{k=0}^{\infty}  f_{k11}(t) z^k \right). 
	\label{FuncDef1}
\end{align}
It is convenient to rearrange \eqref{FuncDef1} to even and odd powers of $z$. 
Introducing the variables $ y:= z^2 $ of the $\Z2$-degree $(0,0)$, one may write  
\begin{align}
	\Phi(t,z,\theta_{10},\theta_{01}) & =\varphi_{00}(t,y)+z\varphi_{11}(t,y)+\theta_{10}(i\psi_{10}(t,y)+z\lambda_{01}(t,y))
	\nn \\
	&+\theta_{01}(i\psi_{01}(t,y)+z\lambda_{10}(t,y))
	+\theta_{10}\theta_{01}(A_{11}(t,y)+zA_{00}(t,y))
	 \label{SupFld}
\end{align}
where
\begin{equation}
	\varphi_{00}(t,y) := \sum_{k=0}^{\infty} f_{2k\,00}(t) y^k, \qquad \varphi_{11}(t,y) := \sum_{k=0}^{\infty} f_{2k+1\,00}(t) y^k, \quad \mathrm{etc.}
\end{equation}
Note that $y$ commutes with all the coordinates. 
The eight component fields $\varphi_{\vec{a}}, \psi_{\vec{a}}, \lambda_{\vec{a}}, A_{\vec{a}}$ are functions of two degree $(0,0)$ variables and their suffixes indicate their respective $\Z2$-degrees ($\deg \Phi = (0,0)$ is assumed). By taking into account the  reality condition \eqref{reality}, the reality condition
\begin{equation}
	\Phi(t,z,\theta_{10},\theta_{01})^{\ast} = \Phi(t,z,\theta_{10},\theta_{01})
\end{equation}
implies that the eight component fields entering \eqref{SupFld} are all real. 

By normalizing the scaling dimension to be $[H]=1$, it follows from \eqref{SUSYalg} and \eqref{Gelement} that
\begin{equation}
	[H]=[Z]=1, \quad [Q_{10}]=[Q_{01}]=\frac{1}{2}, \quad  [t]=[z]=-1,\quad [\theta_{10}]=[\theta_{01}]=-\frac{1}{2}.
\end{equation}
Assuming that $ [\Phi] = s$, the scaling dimensions of the component fields are given by
\begin{alignat}{2}
	s & \quad : & \varphi_{00}
	\nonumber \\
	s+\frac{1}{2} & \quad : & \qquad  \psi_{10}, \ \psi_{01}
	\nonumber \\
	s+1 & \quad : & \varphi_{11}, \ A_{11}
	\nonumber \\
	s+\frac{3}{2} & \quad : & \lambda_{10}, \ \lambda_{01}
	\nonumber \\
	s+2 & \quad : & A_{00}
\end{alignat}

Finally, let us recall the definition of integration of a function $\Phi(X)$ over the minimal $\Z2$-superspace proposed in \cite{NARI}: 
	\begin{equation}
	\int dt\,dz \partial_{\theta_{10}} \partial_{\theta_{01}} \Phi(X) = \frac{1}{2} \int_{D \subset \mathbb{R}^2} dt \,dy\,  A_{00}(t,y) \label{Z22intDef}
\end{equation}
As the notation indicates, integration on the nilpotent variables $\theta_{10}, \theta_{01} $ is the same as the integration of Grassmann variables, namely, integral is identical to derivative. While, the integration on $z$ is converted to integration on $y=z^2$ and the value of integration on $(t,z)$ is evaluated by the ordinary integral on some domain $ D \subset \mathbb{R}^2.$ The domain will be determined by the requirement that $D$ is a support of all the component fields in which $y$ is treated as a real number. 
This is well-defined integral which means that the value of integral remains unchanged under any change of coordinates. 

%%%%%%%%%%%%%%%%%%%%%%%%%%%%%%%%%%%%%%%%%%%%%%%%%%%%%%%%%%%%%%%%%%%%%%
%
\section{$\Z2$-Superfield formalism} \label{SEC:SupefieldFromalization}
\setcounter{equation}{0}

In this section, we develop a $\Z2$-extension of the superfield formalism. Namely, by mimicking the way developed in the standard supersymmetric theory we try to construct a classical action invariant under the transformations generated by the $\Z2$-graded Lie superalgebra $ \g = \mathrm{lin.\ span}\langle \ H, Z, Q_{10}, Q_{01}, L_{11} \ \rangle.$ 

\subsection{Construction of invariant action} \label{SubSEC:construction}

We define the $\Z2$-superfield as a function which is invariant under the transformations generated by $ \g:$
\begin{equation}
	\Phi'(X') = \Phi(X), \qquad X' = X + \delta X
\end{equation}
where $\delta X $ is given in \eqref{CoordinateSUSYTransf} and \eqref{CoordinateLorentzTransf}. 
The transformations of the component fields are determined from 
\begin{align}
	\delta \Phi =& \Phi^\prime(X)  - \Phi(X) \\
	=& (i\epsilon_{00} H+i\epsilon_{11}Z-\epsilon_{10}Q_{10}-\epsilon_{01}Q_{01} +i\epsilon_L L_{11})\, \Phi(X). 
\end{align}
With the notations
\begin{equation}
	\dot{f}(t,y) := \partial_t f(t,y), \qquad 
	f'(t,y) := \partial_y f(t,y)
\end{equation}
they are given as follows:

\medskip\noindent
(i) transformations by $H$
\begin{equation}
	\delta_{00} f = -\epsilon_{00}\dot{f} \qquad \text{for any component fields} \label{SUSYTransH}
\end{equation}
(ii) transformations by $Z$
\begin{align}
	\delta_{11} \varphi_{00}=&-\epsilon_{11}(\varphi_{11}+2y\varphi_{11}^\prime)
	,&
	\delta_{11} \varphi_{11}=&-2\epsilon_{11}\varphi_{00}^\prime
	,\nn \\ 
	\delta_{11} \psi_{10}=& i\epsilon_{11} (\lambda_{01}+2y\lambda_{01}^\prime)
	,&
	\delta_{11} \lambda_{01}=&-2i\epsilon_{11}\psi_{10}^\prime
	,\nn \\
	\delta_{11} \psi_{01}=&i\epsilon_{11} (\lambda_{10}+2y\lambda_{10}^\prime)
	,&
	\delta_{11} \lambda_{10}=&-2i\epsilon_{11}\psi_{01}^\prime
	,\nn \\
	\delta_{11}  A_{11} =&-\epsilon_{11}( A_{00}+2y A_{00}^\prime)
	,&
	\delta_{11}  A_{00} =&-2\epsilon_{11} A_{11}^\prime
	. \label{SUSYTransZ}
\end{align}
(iii) transformations by $Q_{10}$
\begin{align}
	\delta_{10} \varphi_{00}=&-i\epsilon_{10}\psi_{10}
	,&
	\delta_{10} \varphi_{11}=&\epsilon_{10}\lambda_{01}
	, \nn \\
	\delta_{10} \psi_{10}=& \epsilon_{10} \dot\varphi_{00}
	,&
	\delta_{10} \lambda_{01}=&-i\epsilon_{10}\dot{\varphi}_{11}
	,\nn \\
	\delta_{10} \psi_{01}=&i\epsilon_{10} \left(
	A_{11}+\frac{1}{2}\varphi_{11}+y\varphi_{11}^\prime
	\right)
	,&
	\delta_{10} \lambda_{10}=&\epsilon_{10}\left(
	A_{00}+\varphi_{00}^\prime
	\right)
	,\nn \\
	\delta_{10}  A_{11} =&-\epsilon_{10}\left(
	\dot\psi_{01}+\frac{1}{2}\lambda_{01}+y\lambda_{01}^\prime
	\right)
	,&
	\delta_{10}  A_{00} =&-i\epsilon_{10}\left(
	\dot{\lambda}_{10}-\psi_{10}^\prime
	\right)
	. \label{SUSYTransf10}
\end{align}
(iv) transformations by $Q_{01}$
\begin{align}
	\delta_{01} \varphi_{00}=&-i\epsilon_{01}\psi_{01}
	,&
	\delta_{01} \varphi_{11}=&\epsilon_{01}\lambda_{10}
	,\nn\\
	\delta_{01} \psi_{10}=&i\epsilon_{01} \left(
	A_{11}-\frac{1}{2}\varphi_{11}-y\varphi_{11}^\prime
	\right)
	,&
	\delta_{01} \lambda_{01}=&\epsilon_{01}\left(
	A_{00}-\varphi_{00}^\prime
	\right)
	,\nn\\
	\delta_{01} \psi_{01}=& \epsilon_{01} \dot\varphi_{00}
	,&
	\delta_{01} \lambda_{10}=&-i\epsilon_{01}\dot{\varphi}_{11}
	,\nn\\
	\delta_{01}  A_{11} =&-\epsilon_{01}\left(
	\dot\psi_{10}-\frac{1}{2}\lambda_{10}-y\lambda_{10}^\prime
	\right)
	,&
	\delta_{01}  A_{00} =&-i\epsilon_{01}\left(
	\dot{\lambda}_{01}+\psi_{01}^\prime
	\right)
	. \label{SUSYTransf01}
\end{align}
(v) transformations by $L_{11}$
\begin{align}
	\delta_L \varphi_{00} &= \epsilon_L \Big( 2y \dot{\varphi}_{11} + \frac{t}{2}(\varphi_{11}+2y\varphi_{11}') \Big), 
	\nonumber \\
	\delta_L \varphi_{11} &= \epsilon_L (2\dot{\varphi}_{00} + t \varphi_{00}'),
	\nonumber \\
	\delta_L A_{00} &= \epsilon_L (2\dot{A}_{11} + t A_{11}'),
	\nonumber \\
	\delta_L A_{11} &= \epsilon_L \Big( 2y \dot{A}_{00} + \frac{t}{2}(A_{00}+2y A_{00}') \Big),
	\nonumber \\
	\delta_L \psi_{10} &= -i\epsilon_L \Big( 2y \dot{\lambda}_{01} + \frac{t}{2}(\lambda_{01} + 2y\lambda_{01}') -\frac{1}{2} \psi_{01}\Big),
	\nonumber \\
	\delta_L \lambda_{10} &= i\epsilon_L \Big( 2\dot{\psi}_{01} + t\psi_{01}' - \frac{1}{2}\lambda_{01} \Big),
	\nonumber \\
	\delta_L \psi_{01} &= -i\epsilon_L \Big( 2y \dot{\lambda}_{10} + \frac{t}{2}(\lambda_{10} + 2y\lambda_{10}') +\frac{1}{2} \psi_{10}\Big),
	\nonumber \\
	\delta_L \lambda_{01} &= i\epsilon_L \Big( 2\dot{\psi}_{10} + t\psi_{10}' + \frac{1}{2}\lambda_{10} \Big).
	\label{Z22Lorentz}
\end{align}

We introduce the covariant derivatives for the $\Z2$-SUSY transformation \eqref{SUSYTransH}--\eqref{SUSYTransf01}
\begin{equation}
	D_{10} = \partial_{10} - i\theta_{10} \partial_t - \frac{1}{2} \theta_{01} \partial_z, 
	\qquad 
	D_{01} = \partial_{01} - i\theta_{01} \partial_t + \frac{1}{2} \theta_{10} \partial_z
\end{equation}
which satisfy
\begin{alignat}{2}
	\{ D_{\vec{a}}, D_{\vec{a}} \} &= -2H, & \qquad [D_{10}, D_{01}] &= -iZ,
	\nn \\ 
	\{D_{\vec{a}}, Q_{\vec{a}}\} &= 0, & [D_{10}, Q_{01}] &= [D_{01}, Q_{10}] = 0, \quad \vec{a} \in \Z2
\end{alignat}
Note that $D_{\vec{a}}$ is not covariant under $L_{11}$-transformation:
\begin{equation}
	\{ L_{11}, D_{10}  \} = -\frac{1}{2}D_{01}, \qquad \{ L_{11}, D_{01}  \} = \frac{1}{2}D_{10}.
\end{equation}

With these settings, one may write down an action integral which is invariant under the transformations generated by $\g:$
\begin{equation}
	S = -\int dt\, dz \,\partial_{10}  \partial_{01} \, \frac{z}{\sqrt{y}} \Big( 2D_{10}  \Phi \cdot D_{01} \Phi + \alpha V(\Phi) \Big)
	\label{Z22Action}
\end{equation}
where $ \alpha $ is the degree (1,1) coupling constant and $V(\Phi)$ is an arbitrary degree $(0,0) $ function of the superfield. 
The invariance of $S$ is guaranteed by the following observations: 
(i) the integral measure is invariant under \eqref{CoordinateSUSYTransf} and \eqref{CoordinateLorentzTransf} as the $\Z2$-Berezinian for these transformations is 1 \cite{Pz2nint,PonSch} (see also \cite{NARI} for an explicit formula of $\Z2$-Berezinian). 

\noindent
(ii) $z y^{-1/2}$ is invariant under any change of $z.$ 
A change of $ z \to z + \delta z$ induces $y\ \to\ y + \delta y = y + 2z \delta z$, it then follows that
\begin{align}
	\delta \left( \frac{z}{\sqrt{y}} \right) &= \frac{\delta z}{\sqrt{y}} - \frac{z\delta y}{2y^{3/2}} = 0.
\end{align}
(iii) $ D_{10} \Phi $ and $ D_{01}\Phi $ are not covariant under the $L_{11}$-transformation, however, the product $ D_{10} \Phi \cdot D_{01}\Phi $ is covariant due to the nilpotency $ (D_{10} \Phi)^2 = (D_{01} \Phi)^2 = 0:$ 
\begin{align}
	\delta_L (D_{10} \Phi \cdot D_{01}\Phi) &\sim D_{10} L_{11}\Phi \cdot D_{01} \Phi + D_{10}\Phi \cdot D_{01} L_{11} \Phi
	\nonumber \\
	&= \Big( -L_{11} D_{10} \Phi -\frac{1}{2}D_{01}\Phi \Big) D_{01} \Phi + D_{10} \Phi \Big(-L_{11} D_{01} \Phi +\frac{1}{2}D_{10}\Phi  \Big)
	\nonumber \\
	&= - L_{11} \big( D_{10} \Phi \cdot D_{01}\Phi \big).
\end{align}

Now, we evaluate the integral in \eqref{Z22Action} according to the definition \eqref{Z22intDef}. 
To this end, we have to extract degree $(0,0)$ terms proportional to $\theta_{10} \theta_{01} z$ from the integrand in \eqref{Z22Action}. 
Contributions from  $ D_{10} \Phi \cdot D_{01}\Phi $ are obtained by collecting the degree $(0,0)$ terms proportional to $\theta_{10} \theta_{01}$. They are easily identified as
\begin{align}
	\mathcal{K}(t,y) &= -\dot{\varphi}_{00}^2 + y \varphi_{00}'{}^2 -y \dot{\varphi}_{11}^2 + \frac{1}{4}(\varphi_{11}+2y\varphi_{11}')^2-yA_{00}^2-A_{11}^2
	\nonumber \\
	&-i (\psi_{10} \dot{\psi}_{10} + \psi_{01} \dot{\psi}_{01}) -iy (\lambda_{10} \dot{\lambda}_{10} + \lambda_{01} \dot{\lambda}_{01}) + \frac{i}{2}(\psi_{10} \lambda_{10} - \psi_{01} \lambda_{01} ) 
	\nonumber \\
	&-i y (\psi_{10}' \lambda_{10} - \psi_{10} \lambda_{10}' + \psi_{01} \lambda_{01}' - \psi_{01}' \lambda_{01}). 
\end{align}
Contributions from  $ \alpha V(\Phi) $ are computed as follows. We expand $V(\Phi)$ in a power series of  $\theta_{10}, \theta_{01}$ and extract the terms proportional to $\theta_{10}\theta_{01}:$
\begin{align}
	\partial_{10} \partial_{01} V(\Phi)\Big|_{\theta_{10}=\theta_{01}=0}
	&= (A_{11}+zA_{00}) \frac{dV(\varphi)}{d\varphi} 
	\nonumber \\
	&- \big(\psi_{10} \psi_{01} + y\lambda_{10} \lambda_{01} +iz(\psi_{10} \lambda_{10} + \psi_{01} \lambda_{01}) \big) \frac{d^2V(\varphi)}{d\varphi^2}, \label{Potential}
\end{align}
where $ \varphi := \varphi_{00} + z\varphi_{11}. $ We further expand $V(\varphi)$ in a power series of $z \varphi_{11}$ 
%\begin{equation}
%	V(\varphi) = \sum_{n=0} \frac{ (z\varphi_{11})^n}{n!} \partial_{00}^n V(\varphi_{00}) ,
%	\qquad \partial_{00} := \frac{\partial}{\partial \varphi_{00}}.
%\end{equation}
and separate the even and odd powers
\begin{align}
	V(\varphi) &= \sum_{n=0} \frac{y^n \varphi_{11}^{2n}}{(2n)!} \partial_{00}^{2n} V(\varphi_{00}) 
	+ z\sum_{n=0} \frac{y^n \varphi^{2n+1}}{(2n+1)!} \partial_{00}^{2n+1} V(\varphi_{00}), 
	\quad \partial_{00} := \frac{\partial}{\partial \varphi_{00}}.
\end{align}
Using the identities
\begin{equation}
	\frac{dV(\varphi)}{d\varphi} = \partial_{00} V(\varphi), 
	\qquad
	\frac{d^2V(\varphi)}{d\varphi^2} = \partial_{00}^2 V(\varphi) 
\end{equation}
one may obtain 
\begin{equation}
	\partial_{10} \partial_{01} V(\Phi)\Big|_{\theta_{10}=\theta_{01}=0} = \mathcal{V}_{11}(t,y) + z\mathcal{V}_{00}(t,y)
\end{equation}
with
\begin{align}
	&\mathcal{V}_{11} :=A_{11} \tilde{V}_{00} - (\psi_{10} \psi_{01} + y\lambda_{10} \lambda_{01} )\partial_{00}\tilde{V}_{00}
	+ y\Big(A_{00} \tilde{V}_{11}- i (\psi_{10} \lambda_{10} + \psi_{01}\lambda_{01}) \partial_{00}\tilde{V}_{11}\Big),
	\nonumber \\
	&\mathcal{V}_{00} :=  A_{00} \tilde{V}_{00} - i (\psi_{10} \lambda_{10} + \psi_{01}\lambda_{01})\partial_{00}\tilde{V}_{00} + \big( A_{11}\tilde{V}_{11} -(\psi_{10} \psi_{01} + y\lambda_{10} \lambda_{01} )\partial_{00}\tilde{V}_{00} \big)
\end{align}
where
\begin{align}
	\tilde{V}_{00}(t,y) &:= \sum_{n=0} \frac{y^n \varphi_{11}^{2n}}{(2n)!}\, \partial_{00}^{2n+1} V(\varphi_{00}),
	\nn \\
	\tilde{V}_{11}(t,y) &:= \sum_{n=0} \frac{y^n \varphi_{11}^{2n+1}}{(2n+1)!}\, \partial_{00}^{2n+2} V(\varphi_{00}).
\end{align}
Therefore, by the definition \eqref{Z22intDef} we see that only $\mathcal{V}_{11}$ contributes to the integral \eqref{Z22Action}. 

Collecting the contributions, the integral of \eqref{Z22Action} is evaluated as follows:
\begin{equation}
	S = -\int dt\,dy \frac{1}{\sqrt{y}} \left( \mathcal{K} + \frac{\alpha}{2}\mathcal{V}_{11} \right).
\end{equation}
Introducing the degree $(0,0)$ variable $ x := \sqrt{y} $ having the same scaling dimension as $t$ and making the redefinition of variables
\begin{equation}
	\left(\frac{t}{2}, x \varphi_{11}, xA_{00}, x\lambda_{10}, x\lambda_{01}  \right) \ \to \ (t, \varphi_{11}, A_{00}, \lambda_{10}, \lambda_{01}) \label{RedefComponents}
\end{equation}
we obtain a $\Z2$-SUSY Lagrangian with additional $L_{11}$-Lorentz invariance 
\begin{align}
	S &= \int dt\, dx\, (\mathcal{L}_{kin} + \mathcal{L}_{int}),
	\nn \\
	\mathcal{L}_{kin} &:=  \frac{1}{2} (\dot{\varphi}^2_{00}  - \varphi_{00}'{}^2 + \dot{\varphi}^2_{11}  - \varphi_{11}'{}^2) 
	+2A_{00}^2 + 2A_{11}^2
	\nonumber \\
    &
	+i (\psi_{10} \dot{\psi}_{10} + \psi_{01} \dot{\psi}_{01} + \lambda_{10} \dot{\lambda}_{10} + \lambda_{01} \dot{\lambda}_{01})
	\nonumber \\
	& -i(\psi_{10} \lambda_{10}' - \psi_{10}' \lambda_{10} -\psi_{01} \lambda_{01}' + \psi_{01}' \lambda_{01}),
	\nonumber \\
    \mathcal{L}_{int} &:= -2\alpha \big( A_{11} V_{00} + A_{00} V_{11}\big) 
    \nn \\
    &+2\alpha \big(  (\psi_{10} \psi_{01} + \lambda_{10} \lambda_{01} ) \partial_{00}V_{00}
    + i (\psi_{10} \lambda_{10} + \psi_{01}\lambda_{01}) \partial_{00}V_{11} \big)
    \label{Lint}
\end{align}
with
\begin{align}
	V_{00}(t,x) &:= \sum_{n=0} \frac{ \varphi_{11}^{2n}}{(2n)!}\, \partial_{00}^{2n+1} V(\varphi_{00}),
	\nn \\
	V_{11}(t,x) &:= \sum_{n=0} \frac{ \varphi_{11}^{2n+1}}{(2n+1)!}\, \partial_{00}^{2n+2} V(\varphi_{00}).
\end{align}
$V_{00}$ and $ V_{11}$ satisfy the relations
\begin{equation}
	\partial_{00} V_{00} = \partial_{11} V_{11}, \qquad 
	\partial_{11} V_{00} = \partial_{00} V_{11}, \qquad \partial_{11} := \frac{\partial}{\partial \varphi_{11}}.
	  \label{ConstraintV}
\end{equation}
The Lagrangian \eqref{Lint} contains a very general interaction term, the functions $V_{00}, V_{11} $ are chosen almost arbitrary provided that the constraint \eqref{ConstraintV} is satisfied. 
The reason for this is that the derivation of \eqref{Lint} is almost parallel to the derivation of worldline supersymmetry (recall that the minimal $\Z2$-superspace has only one degree $(0,0)$ time coordinate $t$) where we are able to obtain a model with arbitrary potential.    

Despite the fact that the minimal $\Z2$-superspace has only one degree $(0,0)$ coordinate, the Lagrangian obtained is defined in two-dimensional spacetime. This is due to the definition of the integration \eqref{Z22intDef}. 
Namely, the degree $(1,1)$ coordinate $z$ has been converted to the degree $(0,0)$ one in the integration, then we introduced $x$ of degree $(0,0)$ as the square root of $y.$ In this way, the exotic coordinate $z$ is converted to the real number $x$. 

The fields $ A_{00} $ and $ A_{11} $ are auxiliary as seen from \eqref{Lint}. They will be eliminated by the equations of motion. Before doing so, we give the transformation law  of the redefined component fields.  
The $\Z2$-SUSY and $L_{11}$-Lorentz transformations for the redefined variables are readily obtained from \eqref{SUSYTransH}--\eqref{Z22Lorentz}. In the following formulae, the prime stands for the derivative with respect to $x,$ i.e., $ f'(t,x)  = \partial_x f(t,x). $ 

\medskip\noindent
(i) transformation by $H$
\begin{equation}
	\delta_{00} f(t,x) = -\frac{\epsilon_{00}}{2} \dot{f}(t,x), \qquad \text{for any component fields} \label{NewVariTrans0}
\end{equation}
(ii) transformation by $Z$
\begin{align}
	\delta_{11} \varphi_{00}=&-\epsilon_{11}\varphi_{11}^\prime
	,&
	\delta_{11} \varphi_{11}=&-\epsilon_{11}\varphi_{00}^\prime
	,\nonumber \\
	\delta_{11} \psi_{10}=& i\epsilon_{11} \lambda_{01}^\prime
	,&
	\delta_{11} \lambda_{01}=&-i\epsilon_{11}\psi_{10}^\prime
	,\nonumber \\
	\delta_{11} \psi_{01}=&i\epsilon_{11}\lambda_{10}^\prime
	,&
	\delta_{11} \lambda_{10}=&-i\epsilon_{11}\psi_{01}^\prime
	,\nonumber \\
	\delta_{11}  A_{11} =&-\epsilon_{11} A_{00}^\prime
	,&
	\delta_{11}  A_{00} =&-\epsilon_{11} A_{11}^\prime, 
	\label{NewVariTrans1}
\end{align}
(iii) transformation by $Q_{10}$
\begin{align}
	\delta_{10} \varphi_{00}=&-i\epsilon_{10}\psi_{10}
	,&
	\delta_{10} \varphi_{11}=&\epsilon_{10}\lambda_{01}
	,\nonumber \\
	\delta_{10} \psi_{10}=& \frac{1}{2}\epsilon_{10} \dot\varphi_{00}
	,&
	\delta_{10} \lambda_{01}=&-\frac{i}{2}\epsilon_{10}\dot{\varphi}_{11}
	,\nonumber \\
	\delta_{10} \psi_{01}=&i\epsilon_{10} \left(A_{11} + \frac{1}{2}\varphi_{11}^\prime
	\right)
	,&
	\delta_{10} \lambda_{10}=&\epsilon_{10}\left(
	A_{00}+ \frac{1}{2}\varphi_{00}^\prime
	\right)
	,\nonumber \\
	\delta_{10}  A_{11} =&-\frac{1}{2}\epsilon_{10}\left(
	\dot\psi_{01}+\lambda_{01}^\prime
	\right)
	,&
	\delta_{10}  A_{00} =&-\frac{i}{2}\epsilon_{10}\left(
	\dot{\lambda}_{10}-\psi_{10}^\prime
	\right),  \label{NewVariTrans2}
\end{align}
(iv) transformation by $Q_{01}$
\begin{align}
	\delta_{01} \varphi_{00}=&-i\epsilon_{01}\psi_{01}
	,&
	\delta_{01} \varphi_{11}=&\epsilon_{01}\lambda_{10}
	,\nonumber \\
	\delta_{01} \psi_{10}=&i\epsilon_{01} \left(
	A_{11}-\frac{1}{2}\varphi_{11}^\prime
	\right)
	,&
	\delta_{01} \lambda_{01}=&\epsilon_{01}\left(
	A_{00}- \frac{1}{2}\varphi_{00}^\prime
	\right)
	,\nonumber \\
	\delta_{01} \psi_{01}=& \frac{1}{2}\epsilon_{01} \dot\varphi_{00}
	,&
	\delta_{01} \lambda_{10}=&-\frac{i}{2}\epsilon_{01}\dot{\varphi}_{11}
	,\nonumber \\
	\delta_{01}  A_{11} =&-\frac{1}{2}\epsilon_{01}\left(
	\dot\psi_{10}-\lambda_{10}^\prime
	\right)
	,&
	\delta_{01}  A_{00} =&-\frac{i}{2}\epsilon_{01}\left(
	\dot{\lambda}_{01}+\psi_{01}^\prime
	\right)
	. \label{NewVariTrans3}
\end{align}
(v) transformation by $L_{11}$
\begin{align}
	\delta_L \varphi_{00} &= \epsilon_L \hat{\mathfrak{L}} \varphi_{11},
	& 
	\delta_L\varphi_{11} &= \epsilon_L \hat{\mathfrak{L}} \varphi_{00},
	\nonumber \\
	\delta_L A_{00} &= \epsilon_L \hat{\mathfrak{L}} A_{11}, & 
	\delta_L A_{11} &= \epsilon_L \hat{\mathfrak{L}} A_{00},
	\nonumber \\
	\delta_L \psi_{10} &= -i\epsilon_L \Big( \hat{\mathfrak{L}} \lambda_{01} - \frac{1}{2} \psi_{01} \Big),
	& \quad 
	\delta_L \lambda_{10} &= i\epsilon_L \Big( \hat{\mathfrak{L}}\psi_{01} - \frac{1}{2}\lambda_{01} \Big),
	\nonumber \\
	\delta_L \psi_{01} &= -i\epsilon_L \Big( \hat{\mathfrak{L}} \lambda_{10} + \frac{1}{2} \psi_{10} \Big),
	& \quad 
	\delta_L \lambda_{01} &= i\epsilon_L \Big( \hat{\mathfrak{L}}\psi_{10} + \frac{1}{2}\lambda_{10} \Big),
	\nonumber \\
	\hat{\mathfrak{L}} &:= x\partial_t + t \partial_x  \label{InduLorentz}
\end{align}
The generators of these transfromations are given in terms of matrices (the so-called $D$-module presentation). 
Let us arrange the component fields in a vector form:
\begin{equation}
	\tilde{\Phi} := (\varphi_{00}, A_{00},  A_{11}, \varphi_{11}, \psi_{10}, \lambda_{10}, \psi_{01}, \lambda_{01})^T.
\end{equation}
Then, the transformations \eqref{NewVariTrans0}--\eqref{InduLorentz} are recasted in the following form

\medskip\noindent
$ \delta_{00} \tilde{\Phi} =  -i \epsilon_{00} \hat{H} \tilde{\Phi}  \ \Rightarrow \ \hat{H} = \frac{i}{2} \partial_t \mathbb{I}_8. $ 

\medskip\noindent
$ \delta_{11} \tilde{\Phi} = -i \epsilon_{11}\hat{Z} \tilde{\Phi}: $
\begin{equation}
	\hat{Z} = \left(
	\begin{array}{cccc|cccc}
		0 & 0 & 0 & -i\partial_x & 0 & 0 & 0 & 0
		\\
		0 & 0 & -i\partial_x & 0 & 0 & 0 & 0 & 0
		\\
		0 & -i\partial_x & 0 & 0 & 0 & 0 & 0 & 0
		\\
		-i\partial_x & 0 & 0 & 0 & 0 & 0 & 0 & 0
		\\ \hline
		0 & 0 & 0 & 0 & 0 & 0 & 0 & -\partial_x
		\\
		0 & 0 & 0 & 0 & 0 & 0 & \partial_x & 0
		\\
		0 & 0 & 0 & 0 & 0 & -\partial_x & 0 & 0
		\\
		0 & 0 & 0 & 0 & \partial_x & 0 & 0 & 0
	\end{array}
	\right).
\end{equation}

\medskip\noindent
$ \delta_{10} \tilde{\Phi} = -i \epsilon_{10}\hat{Q}_{10} \tilde{\Phi}: $
\begin{equation}
	\hat{Q}_{10} = \frac{1}{2}\left(
	\begin{array}{cccc|cccc}
		0 & 0 & 0 & 0 & 2 & 0 & 0 & 0
		\\
		0 & 0 & 0 & 0 & -\partial_x & \partial_t & 0 & 0
		\\
		0 & 0 & 0 & 0 & 0 & 0 & -i\partial_t & -i\partial_x
		\\
		0 & 0 & 0 & 0 & 0 & 0 & 0 & 2i
		\\ \hline
		i\partial_t & 0 & 0 & 0 & 0 & 0 & 0 & 0 
		\\
		i\partial_x & 2i & 0 & 0 & 0 & 0 & 0 & 0 
		\\
		0 & 0 & -2 & -\partial_x & 0 & 0 & 0 & 0 
		\\
		0 & 0 & 0 & \partial_t & 0 & 0 & 0 & 0 
	\end{array}
	\right).
\end{equation}

\medskip\noindent
$ \delta_{01} \tilde{\Phi} = -i \epsilon_{01}\hat{Q}_{01} \tilde{\Phi}: $
\begin{equation}
	\hat{Q}_{01} = \frac{1}{2}\left(
	\begin{array}{cccc|cccc} 
		0 & 0 & 0 & 0 & 0 & 0 & 2 & 0
		\\
		0 & 0 & 0 & 0 & 0 & 0 & \partial_x & \partial_t
		\\
		0 & 0 & 0 & 0 & -i\partial_t & i \partial_x & 0 & 0
		\\
		0 & 0 & 0 & 0 & 0 & 2i & 0 & 0
		\\ \hline
		0 & 0 & -2 & \partial_x & 0 & 0 & 0 & 0 
		\\
		0 & 0 & 0 & \partial_t & 0 & 0 & 0 & 0 
		\\
		i\partial_t & 0 & 0 & 0 & 0 & 0 & 0 & 0
		\\
		-i\partial_x & 2i & 0 & 0 & 0 & 0 & 0 & 0 
	\end{array}
	\right).
\end{equation}

\medskip\noindent
$ L_{11} \tilde{\Phi} = -i \epsilon_{L} \hat{L}_{11} \tilde{\Phi}: $
\begin{equation}
	\hat{L}_{11} = \left(
	\begin{array}{cccc|cccc}
		0 & 0 & 0 & i\hat{\mathfrak{L}} & 0 & 0 & 0 & 0
		\\
		0 & 0 & i\hat{\mathfrak{L}} & 0 & 0 & 0 & 0 & 0
		\\
		0 & i\hat{\mathfrak{L}} & 0 & 0 & 0 & 0 & 0 & 0
		\\
		i\hat{\mathfrak{L}} & 0 & 0 & 0 & 0 & 0 & 0 & 0
		\\ \hline
		0 & 0 & 0 & 0 & 0 & 0 & -\frac{1}{2} & \hat{\mathfrak{L}}
		\\
		0 & 0 & 0 & 0 & 0 & 0 & -\hat{\mathfrak{L}} & \frac{1}{2}
		\\
		0 & 0 & 0 & 0 & \frac{1}{2} & \hat{\mathfrak{L}} & 0 & 0 
		\\
		0 & 0 & 0 & 0 & -\hat{\mathfrak{L}} & -\frac{1}{2} & 0 & 0 
	\end{array}
	\right). \label{L11Dmod}
\end{equation}
It is easy to verify that the matrix generators satisfy the correct algebraic relations \eqref{SUSYalg} and \eqref{SUSY-Lorentz}.

\subsection{Equations of motion and conserved currents} \label{SubSEC:EoM}

We introduce the two-component spinors of degree $(1,0)$ and $(0,1)$:
\begin{equation}
	\Psi_{10} := \begin{pmatrix}
		\psi_{10} \\ \lambda_{10}
	\end{pmatrix},
	\qquad
	\Psi_{01} := \begin{pmatrix}
		\lambda_{01} \\ -\psi_{01}
	\end{pmatrix}  \label{Choice2}
\end{equation}
which transforms under the $L_{11}$-Lorentz transformation as
\begin{equation}
	\delta_L \Psi_{10} = -i\epsilon_L \Big( \hat{\mathfrak{L}} + \frac{1}{2}\sigma_1 \Big) \Psi_{01},
	\qquad
	\delta_L \Psi_{01} = i\epsilon_L \Big( \hat{\mathfrak{L}} + \frac{1}{2}\sigma_1 \Big) \Psi_{10}
\end{equation}
where $\sigma_1$ is the Pauli matrix. As the transformation has the degree $(1,1)$, spinors $\Psi_{10}$ and $ \Psi_{01} $ are interchanged under the transformation. 
We also introduce the two-dimensional Minkowski metric $ \eta = \mathrm{diag}(1,-1)$ and the  Clifford $\gamma$-matrices defined by the relation:
\begin{equation}
	\gamma^{\mu} \gamma^{\nu} + \gamma^{\nu} \gamma^{\mu} = 2 \eta^{\mu\nu}.
\end{equation}
We take the following representation of $\gamma$-matrices
\begin{equation}
	\gamma^0 = 
	\begin{pmatrix}
		1 & 0 \\ 0 & -1
	\end{pmatrix},
	\qquad
	\gamma^1 = 
	\begin{pmatrix}
		0 & -1 \\ 1 & 0
	\end{pmatrix} 
	\label{ourgamma}
\end{equation}
and define the chirality operator by
\begin{equation}
	\gamma^3 = -\gamma^0 \gamma^1 = 
	\begin{pmatrix}
		0 & 1 \\ 1 & 0
	\end{pmatrix}.
\end{equation}
With $ \partial_{\mu} \equiv (\partial_0, \partial_1) := (\partial_t, \partial_x)$, the Lagrangian \eqref{Lint} is written as follows:
\begin{align}
	\mathcal{L} &= \mathcal{L}_{kin} + \mathcal{L}_{int}
	\nonumber \\
	&= \frac{1}{2} (\partial_{\mu} \varphi_{00} \partial^{\mu} \varphi_{00} + \partial_{\mu} \varphi_{11} \partial^{\mu} \varphi_{11}) +2 A_{00}^2 + 2 A_{11}^2
	+i \overline{\Psi}_{10}  \slashed{\partial} \Psi_{10} 
	+i \overline{\Psi}_{01}  \slashed{\partial} \Psi_{01}
	\nonumber \\
	%	&-2\alpha \left[ A_{11} V_{00} + A_{00} V_{11} + \frac{1}{2}(\overline{\Psi}_{10} \gamma^3 \Psi_{01} - \overline{\Psi}_{01}\gamma^3 \Psi_{10}) \partial_{00} V_{00} 
	%	-\frac{i}{2}( \overline{\Psi}_{10} \gamma^3 \Psi_{10} + \overline{\Psi}_{01} \gamma^3 \Psi_{01}) \partial_{00} V_{11}\right.
	%	\nonumber \\
	&-2\alpha ( A_{11} V_{00} + A_{00} V_{11} )
	\nonumber \\
	& - \alpha \big[ (\overline{\Psi}_{10} \gamma^3 \Psi_{01} - \overline{\Psi}_{01}\gamma^3 \Psi_{10}) \partial_{00} V_{00}
	-i( \overline{\Psi}_{10} \gamma^3 \Psi_{10} + \overline{\Psi}_{01} \gamma^3 \Psi_{01}) \partial_{00} V_{11} \big].
	\label{TotalL}
\end{align}

The fields $ A_{00} $ and $ A_{11}$ are not dynamical, i.e., their equations of motion are given by the algebraic equation
\begin{equation}
	A_{00} = \frac{\alpha}{2}V_{11}, \qquad A_{11} = \frac{\alpha}{2}V_{00}.
\end{equation}
Eliminating these auxiliary fields, the Lagrangian \eqref{TotalL} reads
\begin{align}
	\mathcal{L} 
	&= \frac{1}{2} (\partial_{\mu} \varphi_{00} \partial^{\mu} \varphi_{00} + \partial_{\mu} \varphi_{11} \partial^{\mu} \varphi_{11}) 
	+i \overline{\Psi}_{10}  \slashed{\partial} \Psi_{10} 
	+i \overline{\Psi}_{01}  \slashed{\partial} \Psi_{01}
	\nonumber \\
	&-\frac{\alpha^2}{2} ( V_{00}^2 + V_{11}^2 )
	\nonumber \\
	& - \alpha \big[ (\overline{\Psi}_{10} \gamma^3 \Psi_{01} - \overline{\Psi}_{01}\gamma^3 \Psi_{10}) \partial_{00} V_{00}
	-i( \overline{\Psi}_{10} \gamma^3 \Psi_{10} + \overline{\Psi}_{01} \gamma^3 \Psi_{01}) \partial_{00} V_{11} \big].
	\label{TotalL2}
\end{align}
The Euler-Lagrange equations for the dynamical fields are give by
\begin{align}
	&	\partial_{\mu}\partial^{\mu} \varphi_{00} 
	+\frac{\alpha^2}{2} \partial_{00} (V_{00}^2 + V_{11}^2)
	\nonumber \\
	&\quad 
	+\alpha \big[ (\overline{\Psi}_{10} \gamma^3 \Psi_{01} - \overline{\Psi}_{01}\gamma^3 \Psi_{10}) \partial_{00}^2 V_{00}
	-i( \overline{\Psi}_{10} \gamma^3 \Psi_{10} + \overline{\Psi}_{01} \gamma^3 \Psi_{01}) \partial_{00}^2 V_{11} \big] = 0,
	\nonumber \\[4pt]
	& \partial_{\mu}\partial^{\mu} \varphi_{11} 
	+ \alpha^2 \partial_{00} V_{00}V_{11}
	\nonumber \\
	& \quad 
	+\alpha \big[ (\overline{\Psi}_{10} \gamma^3 \Psi_{01} - \overline{\Psi}_{01}\gamma^3 \Psi_{10}) \partial_{00}^2 V_{11}
	-i( \overline{\Psi}_{10} \gamma^3 \Psi_{10} + \overline{\Psi}_{01} \gamma^3 \Psi_{01}) \partial_{00}^2 V_{00} \big] = 0,
	\nn \\[4pt]
	&i\slashed{\partial} \Psi_{10} + \alpha \gamma^3 (\Psi_{01} \partial_{00} V_{00} - i \Psi_{10} \partial_{00} V_{11}) = 0,
    \nonumber \\[4pt]
     &i\slashed{\partial} \Psi_{01} - \alpha \gamma^3 (\Psi_{10} \partial_{00} V_{00} + i \Psi_{01} \partial_{00} V_{11}) = 0
\end{align}
where we apply the relations \eqref{ConstraintV} to arrange the equations of motion.

Now we turn to the computation of conserved currents associated with the symmetries generated by $\g$.  
Using the standard procedure for computing the Noether currents, we can obtain the followings satisfying $ \partial_{\mu} j^{\mu} = 0.$ 

\medskip\noindent
(i) current associated with $H$
\begin{align}
	j_H^0 &=\frac{1}{2}(\dot{\varphi}_{00}^2 + \varphi_{00}'^2 + \dot{\varphi}_{11}^2 + \varphi_{11}'^2) + i(\psi_{10} \lambda_{10}' - \psi_{10}' \lambda_{10} - \psi_{01} \lambda_{01}' + \psi_{01}' \lambda_{01})
	\nn \\
	&+ \frac{\alpha^2}{2} (V_{00}^2+ V_{11}^2) - 2\alpha(\psi_{10}\psi_{01}+ \lambda_{10}\lambda_{01}) \partial_{00} V_{00} 
	\nn \\
	&-2i\alpha (\psi_{10}\lambda_{10}+\psi_{01}\lambda_{01}) \partial_{00} V_{11},
	\nn \\[4pt]
	j_H^1 &=-\dot{\varphi}_{00} \varphi_{00}' -\dot{\varphi}_{11} \varphi_{11}' +i(\dot{\psi}_{10} \lambda_{10} - \psi_{10}\dot{\lambda}_{10} -\dot{\psi}_{01} \lambda_{01} + \psi_{01} \dot{\lambda}_{01}).
\end{align}
(ii) current associated with $Q_{10}$
\begin{align}
	j_{10}^0 &=  \dot{\varphi}_{00} \psi_{10} + \varphi_{00}' \lambda_{10} -i \dot{\varphi}_{11} \lambda_{01} + i \varphi_{11}' \psi_{01} + \alpha V_{11} \lambda_{10}+i \alpha V_{00} \psi_{01},
	\nonumber \\[4pt]
	j_{10}^1 &= -\dot{\varphi}_{00} \lambda_{10} - \varphi_{00}' \psi_{10} -i\dot{\varphi}_{11} \psi_{01} +i \varphi_{11}' \lambda_{01} - \alpha V_{11} \psi_{10} +i \alpha V_{00} \lambda_{01}.\label{10current}
\end{align}
(iii) current associated with $Q_{01}$
\begin{align}
	j_{01}^0 &=  \dot{\varphi}_{00} \psi_{01} - \varphi_{00}' \lambda_{01} -i \dot{\varphi}_{11} \lambda_{10} -i \varphi_{11}' \psi_{10} +\alpha V_{11} \lambda_{01} + i \alpha V_{00} \psi_{10},
	\nonumber \\[4pt]
	j_{01}^1 &=\dot{\varphi}_{00} \lambda_{01} -  \varphi_{00}' \psi_{01} +i \dot{\varphi}_{11} \psi_{10} +i \varphi_{11}' \lambda_{10} + \alpha V_{11} \psi_{01} -i \alpha V_{00} \lambda_{10}.\label{01current}
\end{align}
(iv) current associated with $Z$
\begin{align}
	j_Z^0 &= \dot{\varphi}_{00} \varphi_{11}' + \varphi_{00}' \dot{\varphi}_{11} + \psi_{10}' \lambda_{01} - \psi_{10} \lambda_{01}' + \psi_{01}' \lambda_{10} - \psi_{01} \lambda_{10}',
	\nn \\[4pt]
	j_Z^1 &=-\dot{\varphi}_{00} \dot{\varphi}_{11} - \varphi_{00}' \varphi_{11}' + \psi_{10} \dot{\lambda}_{01} - \dot{\psi}_{10} \lambda_{01} + \psi_{01} \dot{\lambda}_{10} - \dot{\psi}_{01} \lambda_{10}
	\nn \\
	&+ \alpha^2 V_{00} V_{11} -2\alpha (\psi_{10}\psi_{01} + \lambda_{10}\lambda_{01}) \partial_{00} V_{11} 
	\nn \\
	&-2i\alpha (\psi_{10} \lambda_{10} + \psi_{01} \lambda_{01}) \partial_{00} V_{00}.
\end{align}
(v) current associated with $L_{11}$
\begin{align}
	j_L^0 &= -t j_z^0 - x\big( \dot{\varphi}_{00} \dot{\varphi}_{11} + \varphi_{00}' \varphi_{11}' + \psi_{10}\psi_{01}' - \psi_{10}' \psi_{01}-\lambda_{10} \lambda_{01}' + \lambda_{10}'\lambda_{01} 
	\nn \\
	&+\alpha^2 V_{00} V_{11}  -2\alpha(\psi_{10}\psi_{01} + \lambda_{10}\lambda_{01}) \partial_{00}V_{11} -2i\alpha (\psi_{10}\lambda_{10}+\psi_{01}\lambda_{01}) \partial_{00} V_{00} \big),
	\nn \\[4pt]
	j_L^1 &= -t j_Z^1 + x ( \dot{\varphi}_{00} \varphi_{11}' + \varphi_{00}' \dot{\varphi}_{11} + \psi_{10} \dot{\psi}_{01} - \dot{\psi}_{10}\psi_{01} - \lambda_{10} \dot{\lambda}_{01} + \dot{\lambda}_{10} \lambda_{01} ).
\end{align}

%%%%%%%%%%%%%%%%%%%%%%%%%%%%%%%%%%%%%%%%%%%%%%%%%%%%%%%%%%%%%%%%%%%%%%%%%%%%%%%%%%%%%%%
\section{Examples of $\Z2$-SUSY classical mechanics} \label{SEC:examples}
\setcounter{equation}{0}

In the previous section, we have obtained the Lagrangian \eqref{TotalL2} with an almost arbitrary interaction. 
By an appropriate choice of the interaction, we are able to obtain a $\Z2$-SUSY extension of many known models. 
The analysis of such models will be beyond the scope of the present paper, we  present here two simple but non-trivial examples.  

\subsection{$V(\Phi) =\dfrac{1}{2}\Phi^2$ : massive bosons and interacting fermions}

Let us set $ V(\Phi) =\dfrac{1}{2}\Phi^2$, then $ V_{00} = \varphi_{00}, V_{11} = \varphi_{11}. $ Obviously, the constraints \eqref{ConstraintV} are satisfied. 
With the choice of $V(\Phi)$, the Lagrangian \eqref{TotalL2} yields the following:
\begin{align}
	\mathcal{L} =& \frac{1}{2} (\partial_{\mu} \varphi_{00} \partial^{\mu} \varphi_{00} + \partial_{\mu} \varphi_{11} \partial^{\mu} \varphi_{11}) -\frac{\alpha^2}{2}\varphi_{00}^2-\frac{\alpha^2}{2}\varphi_{11}^2
	\nn \\
	&+i \overline{\Psi}_{10}  \slashed{\partial} \Psi_{10} 
	+i \overline{\Psi}_{01}  \slashed{\partial} \Psi_{01}
	- \alpha (\overline{\Psi}_{10} \gamma^3 \Psi_{01} - \overline{\Psi}_{01}\gamma^3 \Psi_{10}).
	\label{L41}
\end{align}
As $\deg(\alpha^2) = (0,0)$, one may interpret that the bosonic fields $ \varphi_{00} $ and $  \varphi_{11}$ acquire the mass $\alpha^2$, while the fermionic fields remain massless. 
The last term of \eqref{L41} is an interaction terms of different types of fermions. 

The Euler-Lagrange equation for the bosons are massive Klein-Gordon equation
\begin{align}
	(\partial^2+m^2)\varphi_{00}=&0,
	& (\partial^2+m^2)\varphi_{11}=&0
\end{align}
and for fermions we have
\begin{align}
	i\slashed{\partial} \Psi_{10} +\alpha \gamma^3 \Psi_{01} &= 0,
	\nonumber \\
	i\slashed{\partial} \Psi_{01} -\alpha \gamma^3 \Psi_{10} &= 0.
\end{align}

\subsection{$V(\Phi) = \cos \Phi$ : a $\Z2$-extension of sine-Gordon model}

Let us set $V(\Phi) = \cos \Phi$, then 
\begin{equation}
	V_{00} = -\sin\varphi_{00} \cos\varphi_{11}, \qquad 
	V_{11} = -\cos\varphi_{00} \sin\varphi_{11}
\end{equation} 
and the constraints \eqref{ConstraintV} are satisfied. 
The Lagrangian \eqref{TotalL2} yields
\begin{align}
	\mathcal{L} 
	&= \frac{1}{2} (\partial_{\mu} \varphi_{00} \partial^{\mu} \varphi_{00} + \partial_{\mu} \varphi_{11} \partial^{\mu} \varphi_{11}) 
	+i \overline{\Psi}_{10}  \slashed{\partial} \Psi_{10} 
	+i \overline{\Psi}_{01}  \slashed{\partial} \Psi_{01}
	\nonumber \\
	&-\frac{\alpha^2}{2} ( \sin^2\varphi_{00} \cos^2\varphi_{11} + \cos^2\varphi_{00} \sin^2\varphi_{11} )
	\nonumber \\
	& + \alpha \big[ (\overline{\Psi}_{10} \gamma^3 \Psi_{01} - \overline{\Psi}_{01}\gamma^3 \Psi_{10}) \cos\varphi_{00} \cos\varphi_{11}
	+i( \overline{\Psi}_{10} \gamma^3 \Psi_{10} + \overline{\Psi}_{01} \gamma^3 \Psi_{01}) \sin\varphi_{00} \sin\varphi_{11} \big].
	\label{TotalSG}
\end{align}
Note that this Lagrangian is invariant under the exchange $\varphi_{00} \leftrightarrow \varphi_{11}. $ 
The equations of motion obtained from \eqref{TotalSG} is given by
\begin{align}
	\ddot{\varphi}_{00} - \varphi_{00}'' &+ \frac{\alpha^2}{2} \sin 2\varphi_{00} \cos 2\varphi_{11} -\alpha (\overline{\Psi}_{10} \gamma^3 \Psi_{01} - \overline{\Psi}_{01} \gamma^3 \Psi_{10})  \sin \varphi_{00} \cos \varphi_{11} 
	\nonumber \\
	& + i \alpha \big(\overline{\Psi}_{10} \gamma^3 \Psi_{10} + \overline{\Psi}_{01} \gamma^3 \Psi_{01}\big) \cos\varphi_{00} \sin \varphi_{11} = 0,
	\\
	\ddot{\varphi}_{11} - \varphi_{11}'' &+\frac{\alpha^2}{2} \cos 2\varphi_{00} \sin 2 \varphi_{11} -\alpha (\overline{\Psi}_{10} \gamma^3 \Psi_{01} - \overline{\Psi}_{01} \gamma^3 \Psi_{10}) \cos\varphi_{00} \sin \varphi_{11}
	\nonumber \\
	&+ i \alpha \big(\overline{\Psi}_{10} \gamma^3 \Psi_{10} + \overline{\Psi}_{01} \gamma^3 \Psi_{01}\big) \sin \varphi_{00} \cos \varphi_{11},
	\\
	i\slashed{\partial} {\Psi}_{10} &+ \alpha \gamma^3 \Psi_{01} \cos\varphi_{00} \cos\varphi_{11} + i\alpha \Psi_{10} \sin\varphi_{00}\sin\varphi_{11} = 0,
	\\
	i\slashed{\partial} \Psi_{01} &-\alpha \gamma^3 \Psi_{10} \cos\varphi_{00} \cos\varphi_{11} + i\alpha \gamma^3 \Psi_{01} \sin\varphi_{00}\sin\varphi_{11} = 0.
\end{align}
Setting $\varphi_{11} = 0 $ and $\Psi_{10} = \Psi_{01} = 0, $ the above system of equations is reduced to the sine-Gordon equation
\begin{equation}
	\ddot{\varphi}_{00} - \varphi_{00}'' + \frac{\alpha^2}{2} \sin 2\varphi_{00} = 0.
\end{equation}
If we set $ \varphi_{00} = 0$, instead of $\varphi_{11} = 0,$ then we have the sine-Gordon equation for $\varphi_{11}: $
\begin{equation}
	\ddot{\varphi}_{11} - \varphi_{11}'' + \frac{\alpha^2}{2} \sin 2\varphi_{11} = 0.
\end{equation}
Therefore, the Lagrangian \eqref{TotalSG} is a $\Z2$-graded extension of the sine-Gordon model. 
The model \eqref{TotalSG} is different from the one introduced in \cite{bruSG} where the system of equations contains infinitely many component fields. While, the present model has a finite number of component fields, one bosonic, one exotic bosonic and two two-component ferimonic fields. 
It has been shown that the $\Z2$-sine-Gordon model in \cite{bruSG} is integrable. 
It is interesting to investigate the integrability of the model \eqref{TotalSG}. 
It will be reported elsewhere.

%%%%%%%%%%%%%%%%%%%%%%%%%%%%%%%%%%%%%%%%%%%%%%%%%%%%%%%%%%%%%%%%%%%%%%%%%%%%%%%%%%%%%%%
\section{Conclusions} \label{SEC:Conclusions}

In the present work, we have shown that the recently introduced integration on the minimal $\Z2$-superspace \cite{NARI} allows us to construct $\Z2$-supersymmetric actions in a way parallel to the standard supersymmetry. 
Due to the integration with respect to the exotic variables, the Lagrangian obtained is defined in two-dimensional spacetime and  is invariant under the additional transformation generated by $L_{11}$ in \eqref{L11Dmod}. 
Moreover, the Lagrangian has an almost arbitrary interaction terms so that by choosing an appropriate interaction, one may obtain a $\Z2$-graded spersymmetric extension of various known models. 
As an illustration, we gave a $\Z2$-version of supersymmetric sine-Gordon model. 
Although the integrability of the models is still open, our Lagrangian will offer the possibility of new class of integrable systems characterized by the $\Z2$-supersymmetry.

The action integral \eqref{Z22Action}, which has two auxiliary fields, is not the only one we are able to construct.  
It will be possible to construct another action integral which has no or more auxiliary fields as was done in \cite{AKTcl}.  
One may also consider sigma models  or the use of superfields with non-trivial $\Z2$-degree. 
All these possibilities can be carried out with the integration introduced in \cite{NARI}.

From these, it may be said that the integration given in \cite{NARI} is legitimate and that integration defined in a similar way  beyond the minimal $\Z2$-superspace should be investigated. 
It is probably straightforward to increase the number of fermionic coordinates corresponding to the extended $\Z2$-supersymmetry. It is important to consider the $\Z2$-supersymmetry of $\mathcal{N}=2$ because  the most fundamental supersymmetry is ${\cal N} = 2$. 
However, the $\Z2$-supersymmetry discussed in this work is ${\cal N} = 1 $ as each odd subspace has only one supercharge. We remark that there has not yet been much work on $\mathcal{N} = 2$ $\Z2$-supersymmetric theories \cite{AAD,AiDoi}.

%%%%%%%%%%%%%%%%%%%%%%%%%%%%%%%%%%%%%%%%%%%%%%%%%%%%%%%%%%%%%%%%%%%%%%%%%%%%%

\section*{Acknowledgments} 

N. A. is supported by JSPS KAKENHI Grant Number JP23K03217. 

%%%%%%%%%%%%%%%%%%%%%%%%%%%%%%%%%%%%%%%%%%%%%%%%%%%%%%%%%%%%%%%%%%%%%%%%%%%%%

\begin{comment}
\section*{Author's contributions}

All authors contributed equally to this work.

\section*{Data availability}

Data sharing is not applicable to this article as no new data were created or analyzed in this study.
\end{comment}

%%%%%%%%%%%%%%%%%%%%%%%%%%%%%%%%%%%%%%%%%%%%%%%%%%%%%%%%%%%%%%%%%%%%%%%%%%%%%
%
%   References
%

\end{document}